\documentclass[fleqn,usenatbib]{mnras}

\usepackage{newtxtext,newtxmath}

\usepackage[T1]{fontenc}
\usepackage{ae,aecompl}
\usepackage{multirow}
\usepackage{tabularx}
\usepackage{supertabular}
\usepackage{longtable}
\usepackage[]{url}
\usepackage{graphicx}	

\usepackage{multirow}
\usepackage[left]{lineno}
\usepackage{booktabs}
\usepackage{gensymb}
\usepackage{xcolor,color}
\usepackage[caption = false]{subfig}

\newcommand{\beq}{\begin{equation}}
\newcommand{\eeq}{\end{equation}}
\newcommand{\bdm}{\begin{displaymath}}
\newcommand{\edm}{\end{displaymath}}

\definecolor{Gray}{gray}{0.9}
\definecolor{orange}{rgb}{0.9,0.5,0}

\graphicspath{{./plots/}}

\title[Comparing inclination dependent analyses of KN transients]{Comparing inclination dependent analyses of kilonova transients}

\author[J.~Heinzel]{
J.~Heinzel$^{1,2}$, 
M.~W.~Coughlin$^2$,
T.~Dietrich$^3$, 
M.~Bulla$^{4}$, 
S.~Antier$^5$, 
\newauthor
N.~Christensen$^{1,6}$, 
D.~A.~Coulter$^7$,
R.~J.~Foley$^7$,
L.~Issa$^{4,8}$, 
and
N.~Khetan$^9$\\
${}^1$ Carleton College, Northfield, MN 55057, USA \\
${}^2$ School of Physics and Astronomy, University of Minnesota, Minneapolis, Minnesota 55455, USA\\
${}^3$ Institute of Physics and Astronomy, University of Potsdam, 14476 Potsdam, Germany \\
${}^4$ Nordita, KTH Royal Institute of Technology and Stockholm University, Roslagstullsbacken 23, SE-106 91 Stockholm, Sweden\\
${}^5$ APC, UMR 7164, 10 rue Alice Domon et Léonie Duquet, 75205 Paris, France \\
${}^6$ Artemis, Universit\'e C\^ote d'Azur, Observatoire C\^ote d'Azur, CNRS, CS 34229, F-06304 Nice Cedex 4, France\\
${}^7$ Department of Astronomy and Astrophysics, University of California, Santa Cruz, CA 95064, USA\\
${}^{8}$Universit\'e Paris-Saclay, ENS Paris-Saclay,  D\'epartement de Physique, 91190, Gif-sur-Yvette, France. \\
${}^9$ Gran Sasso Science Institute (GSSI), I-67100 L'Aquila, Italy\\
}

\begin{document}
\maketitle

\begin{abstract}
The detection of AT2017gfo proved that binary neutron star mergers are progenitors of kilonovae. 
Using a combination of numerical-relativity and radiative-transfer simulations, 
the community has developed sophisticated models for these transients for a wide portion of the expected parameter space. 
Using these simulations and surrogate models made from them, it has been possible to perform Bayesian inference of the observed signals to infer properties of the ejected matter.
It has been pointed out that combining inclination constraints derived from the kilonova with gravitational-wave measurements increases the accuracy with which binary parameters can be measured and allows a more accurate inference of the Hubble Constant. 
In order to not introduce biases, constraints on the inclination angle for AT2017gfo should be insensitive to the employed models.
In this work, we compare different assumptions about the ejecta and radiative reprocesses used by the community and we investigate their impact on the parameter inference.
While most inferred parameters agree, we find disagreement between posteriors for the inclination angle for different geometries that have been used in the literature. 
According to our study, the inclusion of reprocessing of the photons between different ejecta types improves the modeling fits to AT2017gfo and in some cases affects the inferred constraints.
Our study motivates the inclusion of large $\sim$\,1\,mag uncertainties in the kilonova models employed for Bayesian analysis to capture yet unknown systematics, especially when inferring inclination angles, although smaller uncertainties seem appropriate to capture model systematics for other parameters.
We also use this method to impose soft constraints on the ejecta geometry of the kilonova AT2017gfo.
\end{abstract}

\section{Introduction}

Multi-messenger astronomy is driven by the idea that the observation of a single system with multiple messengers yields a more complete picture of the astrophysical processes than individual observational channels could provide.
Indeed, the first combined detection of gravitational waves (GWs), 
GW170817~\citep{AbEA2017b}, and electromagnetic (EM) 
signals, AT2017gfo and GRB170817A~\citep{2017ApJ...848L..21A,ChBe2017,CoBe2017,2017Sci...358.1570D,2017Sci...358.1565E,2017ApJ...848L..25H,2017Sci...358.1579H,KaNa2017,KiFo2017,2017ApJ...848L..20M,2017ApJ...848L..32M,NiBe2017,2017Sci...358.1574S,2017Natur.551...67P,SmCh2017,2017Natur.551...71T,2017PASJ...69..101U}, from the merger of two neutron stars 
has been a proof of the concept that a multi-messenger approach can be 
successful to constrain the supranuclear equation of state (EOS) of matter, e.g.~\citep{RaPe2018,AnGo2017,BaJu2017,MoWe2018,CoDi2018,RaDa2018,CoDi2018b,CaTe2019,DiCo2020} 
or to measure the expansion 
rate of the Universe, e.g., \citep{2017Natur.551...85A,HoNa2018,CoDi2019,CoAn2020,DiCo2020}. 

Kilonovae (KNe)~\citep{LaSc1974,LiPa1998,MeMa2010,RoKa2011,KaMe2017}, transients in the infrared, optical, 
and ultraviolet bands, are triggered by the radioactive decay of r-process elements produced 
in neutron-rich ejecta released during and after the merger process.
In general, not all binary systems are expected to create KNe, e.g., binary neutron star (BNS) systems with too high total masses~\citep{BaBa2013,AgZa2019,KoBo2019,Bauswein:2020xlt} or black hole-neutron star (BHNS) systems with sufficiently large mass-ratio, 
aligned black hole spins, or compact neutron stars~\citep{Pannarale:2010vs,Foucart:2012nc,Kawaguchi:2016ana,Foucart:2018rjc,Kruger:2020gig}.

In the case of AT2017gfo, the KN signal started out as very blue, 
before reddening on a timescale of a few hours to days. 
While there are still a few possible scenarios to explain the initial, early emission signature~\citep{Arcavi:2018mzm}, 
the red component is likely to be characteristic for almost all KNe
originating from the merger of two neutron stars 
as well as from a BHNS merger, see, e.g. the review of \cite{Me2019} and references therein.
Continuous efforts within the theoretical astrophysics community over the last years
allowed the observed bolometric and photometric data to be connected
with theoretical KN models based on full radiative-transfer simulations~\citep[e.g.,][]{2013ApJ...775..113T,KaKy2016,KaMe2017,BuPa2018,Bulla:2019muo} or analytical/semi-analytical KN models~\citep[e.g.,][]{DiUj2017,Perego:2017wtu,KaSh2018}.

Although KNe are likely observable from all directions and not beamed as short gamma-ray bursts, 
numerical relativity simulations~\citep[e.g.,][]{HoKi13,DiUj2017,Radice:2018pdn} indicate that 
the ejected matter in a neutron-star merger is not perfectly spherical.
Thus, not only the photon emission shows a clear angular dependence, also the entire ejecta geometry can vary. 
Broadly speaking, there are at least two geometric sections of the KN: first, the material ejected around the moment of merger via torque and shocks, called \textit{dynamical ejecta}, with light r-process material primarily distributed in the polar regions and heavier tidal r-process material concentrated towards the equatorial plane, \citep[e.g.,][]{Wanajo:2014wha,Kawaguchi:2016ana,DiUj2017}.
The second section is ejected after the merger, by winds produced from the remnant system due to neutrino emission, magnetic fields, or secular effects that drive further ejection~\citep[e.g.,][]{DeOt2009,PeRo2014,Martin:2015hxa,KiSe2015,FeKa2015,KaFe2015,
FoOC2016,SiMe2017,Radice:2018pdn}. This component is called the \textit{disk wind}, and is often approximated as being free of heavier r-process material.

KNe are often simulated by radiative transfer algorithms, which simulate emission and propagation of radiation in r-process material.
Since the distribution of radioactive material is directly related to the distribution of mass, assumptions must be made about the underlying geometry and behavior of the ejected material. 
These geometries are a critical assumption, as they directly lead to inference on parameters such as the inclination or mass of the ejected material, which then are tied back to progenitor parameters.

In this paper, we will use several different KN geometries and the radiative transfer code \texttt{POSSIS} developed by \cite{Bulla:2019muo} to understand how inclination constraints on GW170817/AT2017gfo depend on the underlying geometry assumed, extending the work of \cite{Dhawan:2019phb} and complementary to the work of \cite{KoBa2020}.
In Section~\ref{sec:models}, we discuss the creation of surrogate models based 
on KN geometries inspired by \cite{KaMe2017}, \cite{WoKo2018} and \cite{Bulla:2019muo}.
These surrogate models are direct extensions of~\cite{CoDi2018} and provide a phenomenological description of inclination effects. In Section~\ref{sec:inclination}, we compare the surrogate models 
against each other, in particular investigating possible systematic biases and uncertainties and explore how each of the surrogate models derived in Section~\ref{sec:models} 
provide constraints on the inclination angle of GW170817/AT2017gfo, 
the total ejected mass, and the dynamical-to-disk wind mass ratio.
We conclude in Section~\ref{sec:conclusion} and outline avenues for future work.

\section{KN models and surrogate construction}
\label{sec:models}

In the following, we explore three different models that simulate multiple viewing-angle-dependent components. It has become common to simulate ejecta separately, then add the lightcurves in flux space after simulation~\citep[e.g.,][]{KaMe2017,Villar:2017wcc,CoDi2018}, which neglects any sort of \textit{reprocessing} between multiple components, when radiation from one component is absorbed and re-radiated at a different wavelength by the other component. 
We also examine how accounting for this reprocessing can change the lightcurve fitting to AT2017gfo and the inferred posteriors on the system parameters; cf.~\cite{KaSh2018}.

All models that we investigate are axially symmetric, following the precedent of \cite{KaMe2017,WoKo2018,Bulla:2019muo,KaSh2020,korobkin2020}. Thus, our radiative transfer simulations give us inclination dependent lightcurves, where we compute lightcurves for $11$ different inclination angles $\theta_{\rm obs}$ using a uniform spacing in $\cos \theta_{\rm obs}$.

\subsection{\texttt{POSSIS}}
\label{sec:possis}

\cite{Bulla:2019muo} developed a radiative transfer code \texttt{POSSIS}, in which initially a spherical KN geometry had been employed with  a lanthanide rich component near the equatorial region and a lanthanide free component towards the poles; see the left panel of Fig. \ref{fig:geometries}. Bound-bound opacities are treated as either ``rich'' (electron fraction $Y_e \le 0.25$) or ``free'' ($Y_e > 0.25$) in \texttt{POSSIS}, with their wavelength- and time-dependence chosen to mimic realistic opacities base on atomic calculations from \cite{Tanaka2018}.

\texttt{POSSIS} creates photon packets based on the distribution of radioactive material and emits them isotropically. 
The frequencies of the photon packets produced are sampled from the thermal emissivity at the precise location, and  photon packet energy is determined from nuclear heating rates in \cite{KoRo2012}.

\texttt{POSSIS} propagates the photon packets through the ejecta (assumed to expand homologously) until they interact with matter. It then probabilistically decides whether this interaction is governed by electron scattering, or by a bound-bound, bound-free, or free-free opacity. In electron scattering, the frequency is unchanged (in the interaction frame) and the new direction is sampled from a scattering probability distribution (see Equation 12 of \citealt{Bulla2015}). If a bound-bound transition occurs, \cite{Bulla:2019muo} uses the two-level atom approach (TLA) of \cite{KaTh2006}, where the photon packet is re-emitted 10\% of the time with the same frequency while 90\% of the time with a new frequency sampled from the location's thermal emissivity.
The TLA approach is meant to model the complex behavior of re-emitting a photon through many different possible line transitions (for more detailed investigation on the assumptions inherent to the TLA approach, see Eq. 7 and Section 3.6 of \cite{KaTh2006}). Bound-free and free-free processes are subdominant at the relevant wavelengths for KNe \citep{Tanaka2018} and are thus typically neglected.

We used an improved version of \texttt{POSSIS}, which instead of taking the thermalization efficiency to be $\epsilon_{\rm th} = 0.5$, takes the thermalization efficiencies from \cite{BaKa2016}. Furthermore, temperature is no longer a free parameter, and is estimated self-consistently from the mean intensity of the radiation \citep{CoAn2020,Carracedo:2020xhd}.
 
\subsection{Spherical Segment-Spherical Cap (SSCr) Geometry}
 
\begin{figure*}
\centering
\includegraphics[width=.33\textwidth]{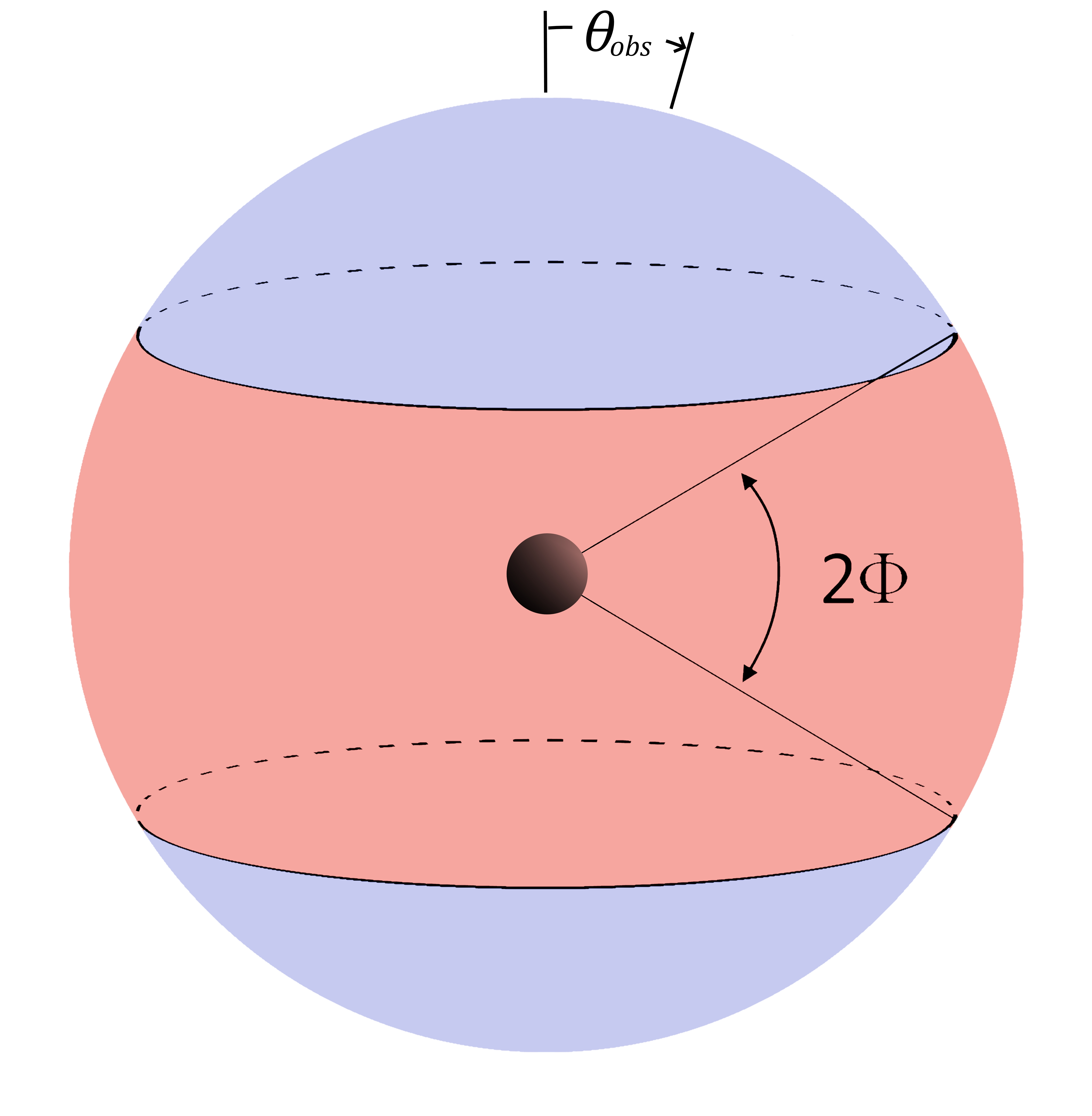}
\includegraphics[width=.33\textwidth]{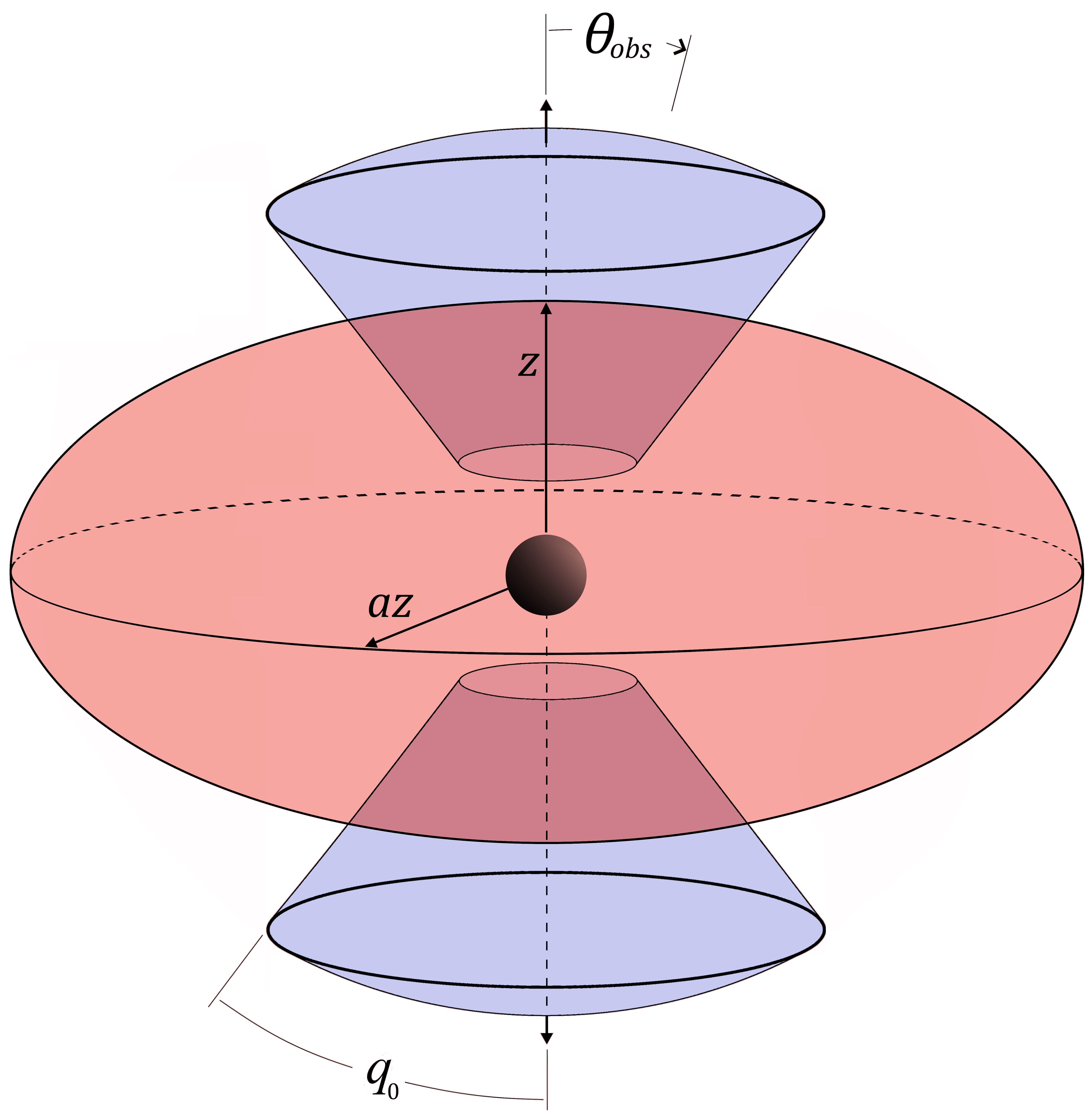}
\includegraphics[width=.33\textwidth]{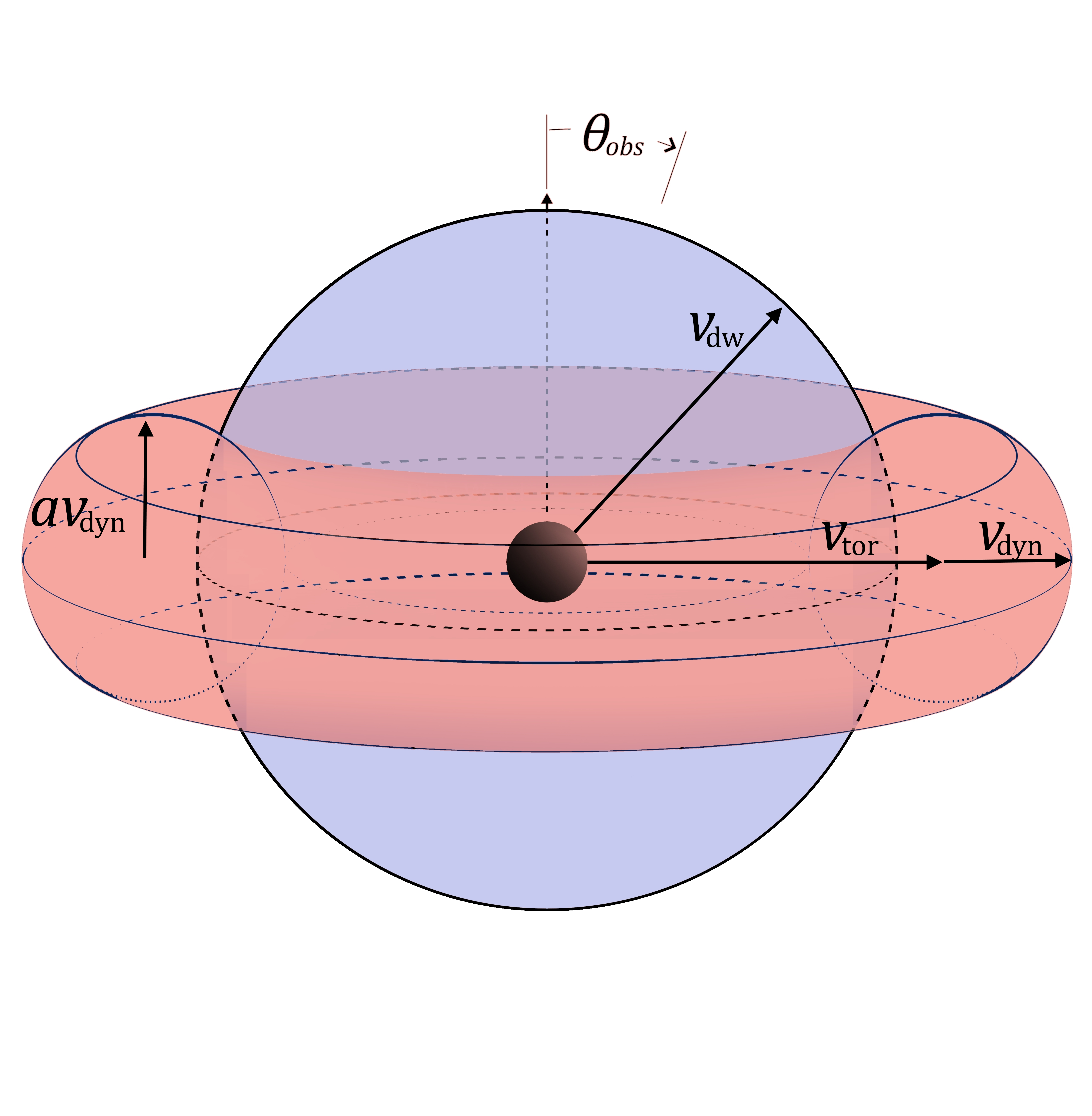}
\caption{Depictions of the ejecta geometries of the SSCr model in the left panel, the ElDC model in the middle panel, and the TorS model in the right panel. 
Blue colors refer to lanthanide poor regions and red refers to lanthanide rich regions. 
On the left, \protect\cite{Bulla:2019muo} defined their half-opening angle $\Phi$  to be analogous to $90^\circ - q_0$, where $q_0$ is the half-opening angle as defined by \protect\cite{KaMe2017}; we use the same definitions.
In the middle panel, we show \protect\cite{KaMe2017}'s ejecta geometries; 
\protect\cite{KaMe2017} simulated aspect ratios $a$ of $a=2$ or $a=4$,
with only the polar dynamical ejecta, and the half-opening angle of the polar dynamical ejecta fixed to be $q_0=45^\circ$.
On the right panel, \protect\cite{WoKo2018} simulated the tidal dynamical ejecta according to direct Numerical Relativity simulations; in our simulations (the TorS model), we adopted a approximation of their technique.}
\label{fig:geometries}
\end{figure*}

Due to the relatively fast analysis speeds - a typical simulation takes a few hours to a day on a single core - \cite{CoAn2020} were able to use \texttt{POSSIS} to simulate their KN model over a full grid of two parameter values for their geometry (Fig. \ref{fig:geometries} left panel) - half opening angle $\Phi$, and total ejecta mass $M_{\rm ej}$. We use the 50 simulations from \cite{CoAn2020}, and call it the Spherical Segment-Spherical Cap reprocessing (SSCr) Geometry, based on its qualitative shape (see Fig. \ref{fig:geometries}). Since reprocessing, i.e. when radiation from one ejecta component is absorbed by the other and reradiated at a different wavelength, between the two components was incorporated in the simulations from \cite{CoAn2020}, we will refer to this geometry as SSCr to be consistent with the other KN models described below.

\subsection{Ellipsoidal-Double Cone (ElDC) Geometry}

\cite{KaMe2017} explored non-spherical geometries by using a two-dimensional axially symmetric geometry, with separate components for the polar dynamical ejecta and the disk wind ejecta.
For the polar ejecta, they used an ad-hoc, broken power-law density profile describing 
the distribution of the ejected material (with mass $M_{\rm ej, 1}$) over varying radii and velocities, 
declining gradually in the interior but dropping steeply in the outer layers
\begin{equation}
\rho_{\rm pol}(r, \theta, t) =  \rho_{\rm pol}(r, t) \left[1 + \left(\frac{f (\theta)}{f_0}\right)^{10}\right]^{-1},
\end{equation}
where $\rho_{\rm pol}(r, t)$ is the normalized spherical density profile, 
shown in Equation 4 of \cite{KaMe2017}, $f(\theta) = 1 - \cos(\theta)$, and $f_0 = 1 - \cos(q_0)$. 
The idea is that the formula concentrates the ejecta in the polar cone with half-opening angle $q_0$. 
The ellipsoid model parameters simulated in that work were fixed to an ejecta mass $M_{\rm ej,2}$ = 0.04\,$M_\odot$, an ejecta velocity $v_{\rm ej}$ = 0.1c, and a lanthanide fraction $X_{\rm lan}$ = $10^{-2}$. The axis ratio and opening angle assumed in their work were $a=4$ and $q_0=45^\circ$, respectively.

In our analysis, we extended these two geometry grids for the disk wind and polar dynamical ejecta formulations, and simulated a grid of geometries with the four parameters $M_{\rm ej, 1}, M_{\rm ej, 2}, q_0$, and $a$,  outlined in Table \ref{tab:kasen_grid}. We refer to this model as the Ellipsoidal-Double Cone (ElDC) model.
We \emph{first} simulate the disk wind and dynamical geometries separately, i.e., we have a ``dynamical'' simulation set and a ``disk wind'' simulation set. As done in e.g. \cite{KaMe2017,Villar:2017wcc,CoDi2018}, we add the lightcurves together, first converting to flux space, and then back into magnitudes. 

\begin{table}
     \centering
     \begin{tabular}{| c | c | p{3.2cm} |}
     \hline
     Ejecta type & Parameter & Parameter Values \\
     \hline
     Dynamical & $M_{\rm ej, 1}$ & 
     $[0.001,$ $0.01,$ $0.025,$ $0.04,$ $0.055,$ $0.07,$ $0.085,$ $0.1]M_\odot$ \\ \hline
     Dynamical & $q_0$ & $[15^\circ,30^\circ,45^\circ,60^\circ,75^\circ]$
     \\ \hline
     Disk wind & $M_{\rm ej, 2}$ & 
     $[0.001,$ $0.01,$ $0.025,$ $0.04,$ $0.055,$ $0.07,$ $0.085,$ $0.1]M_\odot$ \\ \hline
     Disk wind & $a$ & $[1, 2, 4, 6, 8]$ \\ \hline
     \end{tabular}
\caption{Summary of the parameters in the ElDC geometry simulated in \texttt{POSSIS}.}
\label{tab:kasen_grid}
\end{table}

Given the speed of \texttt{POSSIS}, it is possible to combine the geometries to investigate ``reprocessing'' effects, when radiation escaping from one ejecta component is absorbed by the other and re-radiated at the same or at a different frequency (see Section~\ref{sec:possis}). 
To do so, we combined these two geometries together into a single geometry by superimposing each disk wind with each dynamical geometry. 
Rather than superimpose the ejecta directly and mix the lanthanide fractions, we take a different approach. Because of how \texttt{POSSIS} treats lanthanide opacities, we needed to ensure directly that the masses allocated to each component were what we desired. For each cell, the lanthanide fraction is taken to be lanthanide-rich if the density contribution from the dynamical component is higher than the contribution from the disk wind, and lanthanide-free if the opposite is true. 
The total mass of each ejecta is then rescaled to match the desired mass in each ejecta. Since this may mean some cells identified as lanthanide-free should become lanthanide-rich (or vice-versa), the above process is done iteratively until a stable point is found, that is, when there are no more cells that switch from lanthanide-free to lanthanide-rich. If the process gets stuck in a meta-stable loop, the process is terminated (usually only a few cells are affected in this case). 
We refer to the simulation set with reprocessing as the Ellipsoidal-Double Cone Reprocessing (ElDCr) model.

It is important to note that in the typical method, where ejecta geometries are simulated separately, only a grid of $8\times 5 + 8\times 5 = 80$ simulations (see Table \ref{tab:kasen_grid}) were needed to cover necessary parameter space. When multiple geometries are reprocessed, $8\times 5 \times 8 \times 5 = 1600$ simulations were necessary. 
So this is a case where computational efficiency is sacrificed for modeling accuracy.

\subsection{Toroidal-Spherical (TorS) Geometry}

We also employ \texttt{POSSIS} to simulate geometries similar to those presented in \cite{WoKo2018}. \cite{WoKo2018} also used a two-component model, simulating only \textit{tidal} dynamical ejecta directly, using long-term numerical relatively simulations, treated as lanthanide-rich in composition. For the second component, they simulated a spherical disk wind treated as lanthanide-poor, with the density profile
\begin{equation}
    \rho(v) = \rho_0 \left(1-\frac{v^2}{v_{\rm dw}^2} \right)^3,
    \label{eq:Wollaeger_dw_density}
\end{equation}
where $\rho_0$ is a coefficient proportional to the total ejecta mass.
\cite{WoKo2018} fixed $v_{\rm dw} = 0.15$c, whereas in the following, we treat $v_{\rm dw}$ as a free parameter.

\begin{figure}
    \centering
    \includegraphics[width=\columnwidth]{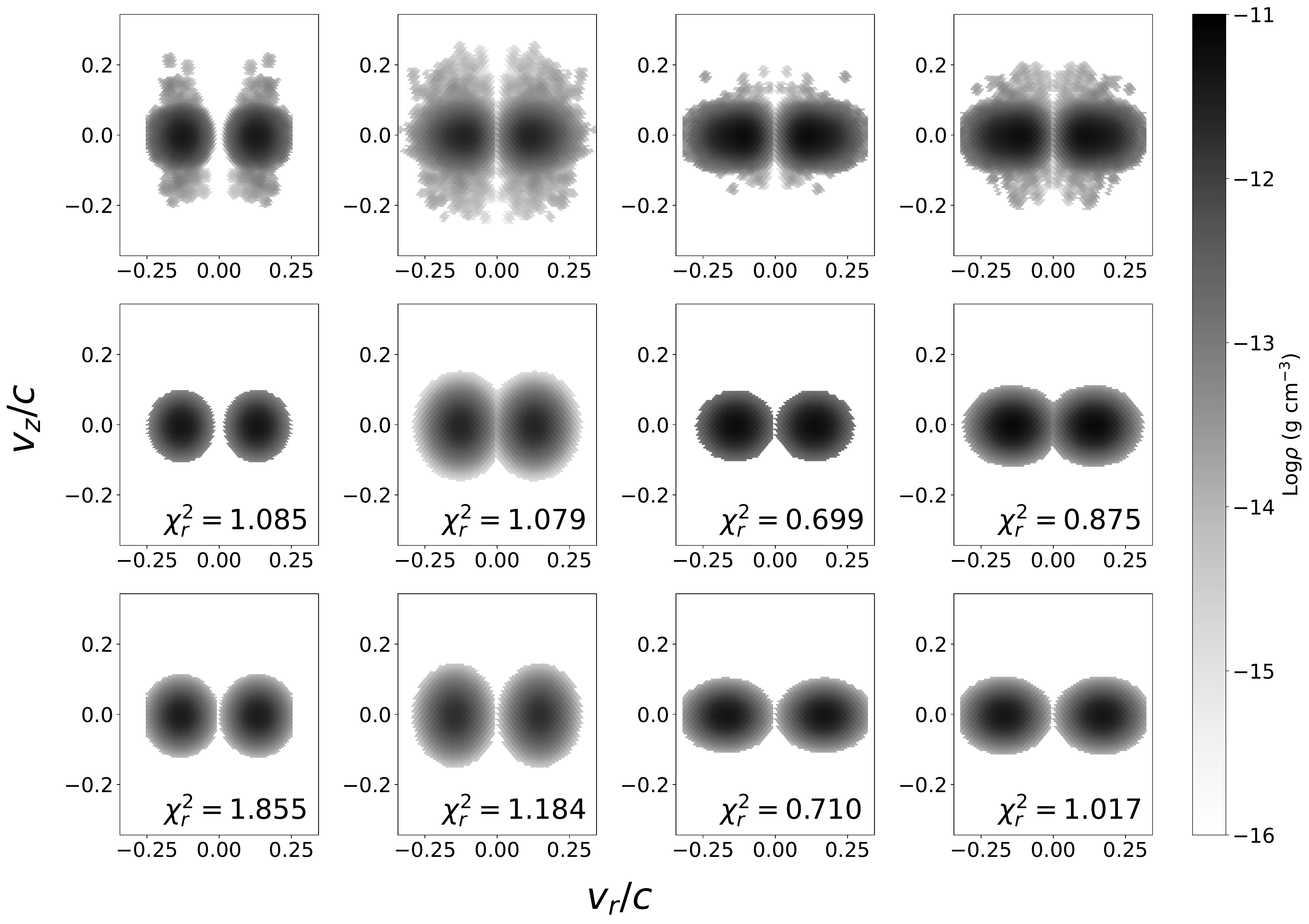}
    \caption{A comparison between the axially averaged  \protect\cite{RoKo2014} models used in  \protect\cite{WoKo2018} (top panels), the $\chi^2$ minimization of Eq. \ref{eq:Wollaeger_dyn} with respect to all five parameters: $\rho_0, v_{\rm inner}, a, \sigma$, and $v_{\rm dyn}$ (middle panels), and the $\chi^2$ minimization of Eq. \ref{eq:Wollaeger_dyn} with $v_{\rm inner}=0.059c+v_{\rm dyn}$ and $\sigma=v_{\rm dyn}/3.5$ fixed (bottom panels). From left to right are the simulations A, B, C, and D from  \protect\cite{RoKo2014}.}
    \label{fig:chi2_min_wollaeger}
\end{figure}

As there are only four long-term numerical relativity (NR) simulations from \cite{RoKo2014}, the simulation set is not large enough for our interpolator to gain meaningful constraints \citep{MacKay1998IntroductionTG}. However, since the general shape of each NR simulation geometry was seen to be toroidal (see Fig. \ref{fig:chi2_min_wollaeger}) in \cite{RoKo2014}, we fit a Gaussian-toroid described by three free parameters: in cylindrical coordinates, it is defined as
\begin{equation}
    \rho(v_r,v_z) =\begin{cases}
    \rho_0 \exp{\left(\frac{-(v_r - v_{\rm inner})^2 + \left(v_z/a\right)^2}{2\sigma^2}\right)} & f(v_r,v_z) \le v_{\rm dyn} \\
    0 & f(v_r,v_z) > v_{\rm dyn}
    \end{cases}
    \label{eq:Wollaeger_dyn}
\end{equation}
where $f(v_r,v_z)=\sqrt{(v_r - v_{\rm inner})^2 + \left(v_z/a\right)^2}$. 
Here, $\rho_0$ is a density coefficient proportional to the ejecta mass, $v_r = \sqrt{v_x^2+v_y^2}$ is the radial velocity, $v_{\rm inner} = 0.059\,{\rm c} + v_{\rm dyn}$ represents the central toroidal radius (with $0.059$c as the fixed inner radius), $a$ is the axis ratio, and $\sigma = v_{\rm dyn}/3.5$ is the standard deviation of the gaussian. 

To minimize the number of dimensions for the Gaussian Process interpolator, we fixed $v_{\rm inner}$ and $\sigma$ to those values that minimized the $\chi^2$ value.
We then simulated a grid of dynamical and disk-wind geometries as in Table~\ref{tab:wollaeger_grid}, and added the lightcurves together post-simulation as done for the ElDC model. 
We present the $\chi^2$-minimization fits in Fig. \ref{fig:chi2_min_wollaeger}. The bottom panels are very similar to the middle panels, with perhaps slightly extended boundaries, and the reduced $\chi^2$'s are only slightly larger. We also simulated each of these geometries with \texttt{POSSIS}, and found no noticeable discrepancy between the lightcurves, further indicating that these fits are appropriate for radiative transfer.

\begin{table}
    \centering
    \begin{tabular}{| c | c | p{3.55cm} |}
     \hline
     Ejecta type & Parameter & Parameter Values \\
     \hline
     Dynamical & $M_{\rm ej, 1}$ & 
     $[0.001,$ $0.01,$ $0.025,$ $0.04,$ $0.055,$ $0.07,$ $0.085,$ $0.1]M_\odot$ \\ \hline
     Dynamical & $a$ & $[0.25,0.5,0.75,1,1.5,2]$
     \\ \hline 
     Dynamical & $v_{\rm dyn}$ & $[0.04,$ $0.085,$ $0.12,$ $0.155,$ $0.2,$ $0.25]$c \\ \hline
     Disk wind & $M_{\rm ej, 2}$ & 
     $[0.001,$ $0.01,$ $0.025,$ $0.04,$ $0.055,$ $0.07,$ $0.085,$ $0.1]M_\odot$ \\ \hline
     Disk wind & $v_{\rm dw}$ & $[0.1,$ 0.15, 0.2, 0.25, 0.3, $0.35]$c \\ \hline
     \end{tabular}
    \caption{Summary of the parameters in the TorS geometry simulated in \texttt{POSSIS}.}
    \label{tab:wollaeger_grid}
\end{table}

As with the ElDC and ElDCr simulation sets, we constructed both a simulation set where the ejecta were processed independently, and a simulation set with ``reprocessing.'' The dynamical grid had $8\times6\times6 = 288$ simulations, and the disk wind grid had $8\times6=48$ simulations. As we will mention in Sec. \ref{sec:inclination}, we reduced the simulation set for ``reprocessing,'' as a full grid would require $288\times48=13824$ simulations, which is more than we can feasibly do. In addition, training the Gaussian Process Interpolator is the bottleneck for our analysis procedure, not producing simulations. Thus, rather than produce a full simulation set, we reduced the number of parameters in each dimension by a factor of 2, so in the end we only created 432 simulations.

We refer to this simulation set and the derived model as the Toroidal-Spherical (TorS) model. With reprocessing accounted for, we call it the Toroidal-Spherical Reprocessing (TorSr) model.


\section{Inclination and Interpretation of GW170817}
\label{sec:inclination}

\begin{figure}
 \includegraphics[width=\columnwidth]{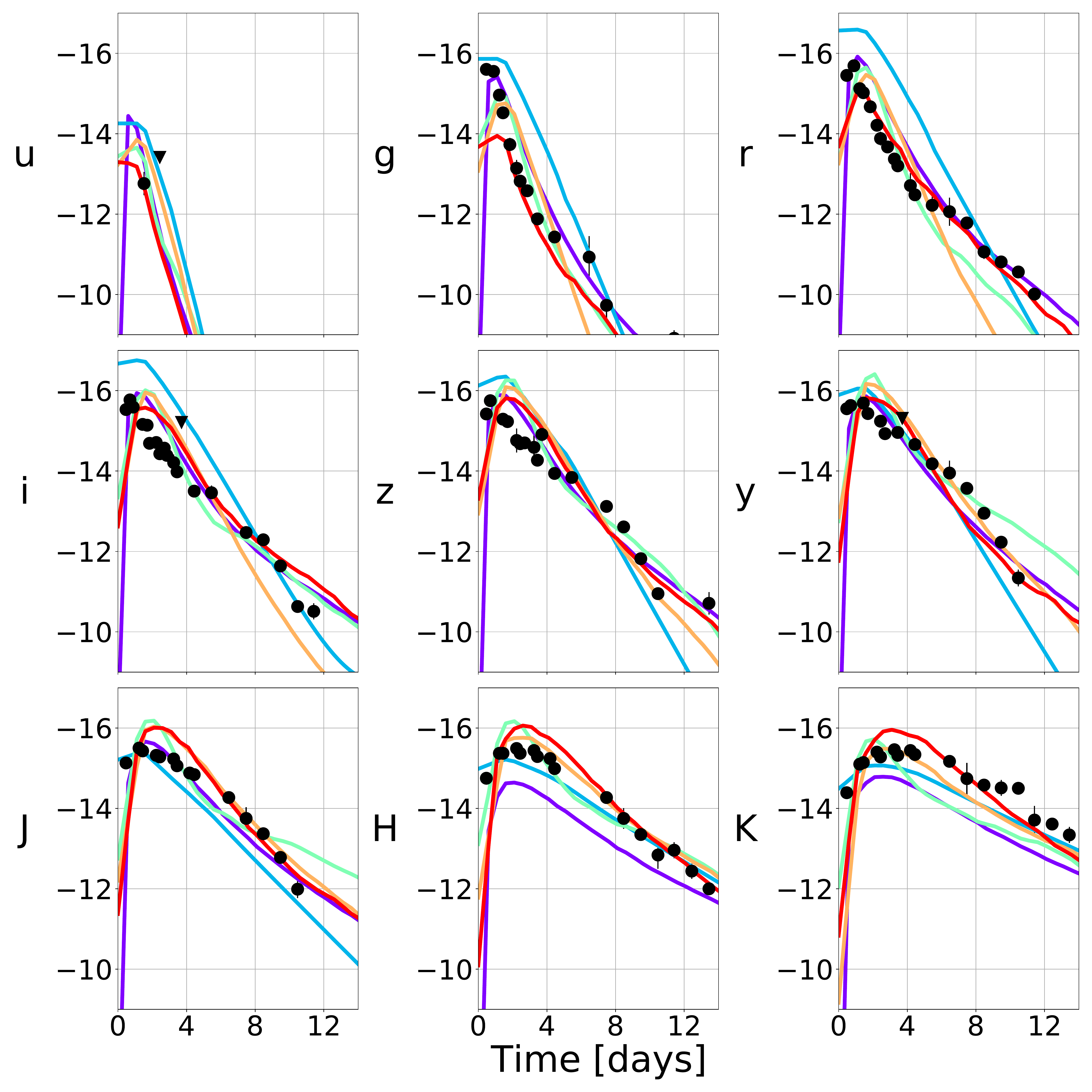}
 \caption{Lightcurves for the inclination-dependent surrogate KN models with the points of GW170817 included (with the data taken from the sample from \protect\cite{CoDi2018b}). The lightcurves shown correspond to a maximum likelihood $\chi^2$ fit to the data for the intrinsic parameters, assuming a $\pm 0.5$ mag uncertainty in the lightcurve fitting, in quadrature with the photometric uncertainty from the data. All the lightcurves are expressed in AB absolute magnitudes. The circles denote actual detections while the triangles are upper limits. The letters to the left of the y-axes show the passbands of the observations. Purple is the SSCr model with a reduced $\chi^2$ of 2.80, light blue is the ElDC model with a reduced $\chi^2$ of 4.18, light green is the ElDCr model with a reduced $\chi^2$ of 1.85, orange is the TorS model with a reduced $\chi^2$ of 3.70, and in red is the TorSr model with a reduced $\chi^2$ of 1.79.}
 \label{fig:GW170817}
\end{figure}

\subsection{Inclination-dependent surrogates}

To use these simulations in the surrogate lightcurve models, we introduce a modification of our previous technique~\citep{CoDi2017,CoDi2018,CoDi2018b}. 
We use the five different grids of radiative transfer simulations described above as training for a Gaussian Process Regression based method. 
This allows us to interpolate the lightcurves across the continuous parameter space of interest.
To do so, we take each simulation and compute $u$, $g$, $r$, $i$, $z$, $y$, $J$, $H$, and $K$-band lightcurves as a function of the viewing angle.
Not only is Gaussian process non-parametric, so the form of our parameterization for each model will not introduce bias into the parameter estimation, but it also is Bayesian in nature; by comparing the inferred lightcurve for a set of parameters to the measured lightcurve for AT2017gfo, we can compute likelihoods based on a given set of model parameters. 
For each of the simulation sets, we assume flat priors (flat in log for the ejecta masses) that span the full parameter space we simulated, but usually do not extend far beyond the simulations. For the SSCr simulation set, we used $-3 \le \log_{10}\left(\frac{M_{\rm ej}}{M_\odot}\right) \le 0$ and $0^\circ \le \Phi \le 90^\circ$; 
in the ElDC and ElDCr simulation sets, we used $0^\circ \le q_0 \le 90^\circ$ and $1 \le a \le 10$, and 
in the TorS and TorSr simulations, we used $0.25 \le a \le 2$, $0.04c \le v_{\rm dyn} \le 0.25c$, and $0.1c \le v_{\rm dw} \le 0.35c$.
For the ElDC, ElDCr, TorS and TorSr simulation sets, we used $-3 \le \log_{10}\left(\frac{M_{\rm ej, 1}}{M_\odot}\right), \log_{10}\left(\frac{M_{\rm ej, 2}}{M_\odot}\right) \le 0$, and finally we used $0^\circ \le \theta_{\rm obs} \le 90^\circ$ for all the simulation sets.
The priors were the same for the reprocessing versus the separately simulated KNe.

For the reprocessing simulation sets, the full set of simulated training data was too large for the Gaussian Process computation to sample effectively, so we instead trained the algorithm on a stratified set, decimated by a factor of 2 in each dimension (except for inclination angle). We determined that the reduced training set provided appropriate constraints by comparing constraints from the stratified sample to those with a smaller sample; these were shown to be indistinguishable, and therefore we expect that adding more training lightcurves would not have led to a noticeable improvement. 

\subsection{Parameter Estimation}

\begin{figure}
    \centering
    \includegraphics[width=\columnwidth]{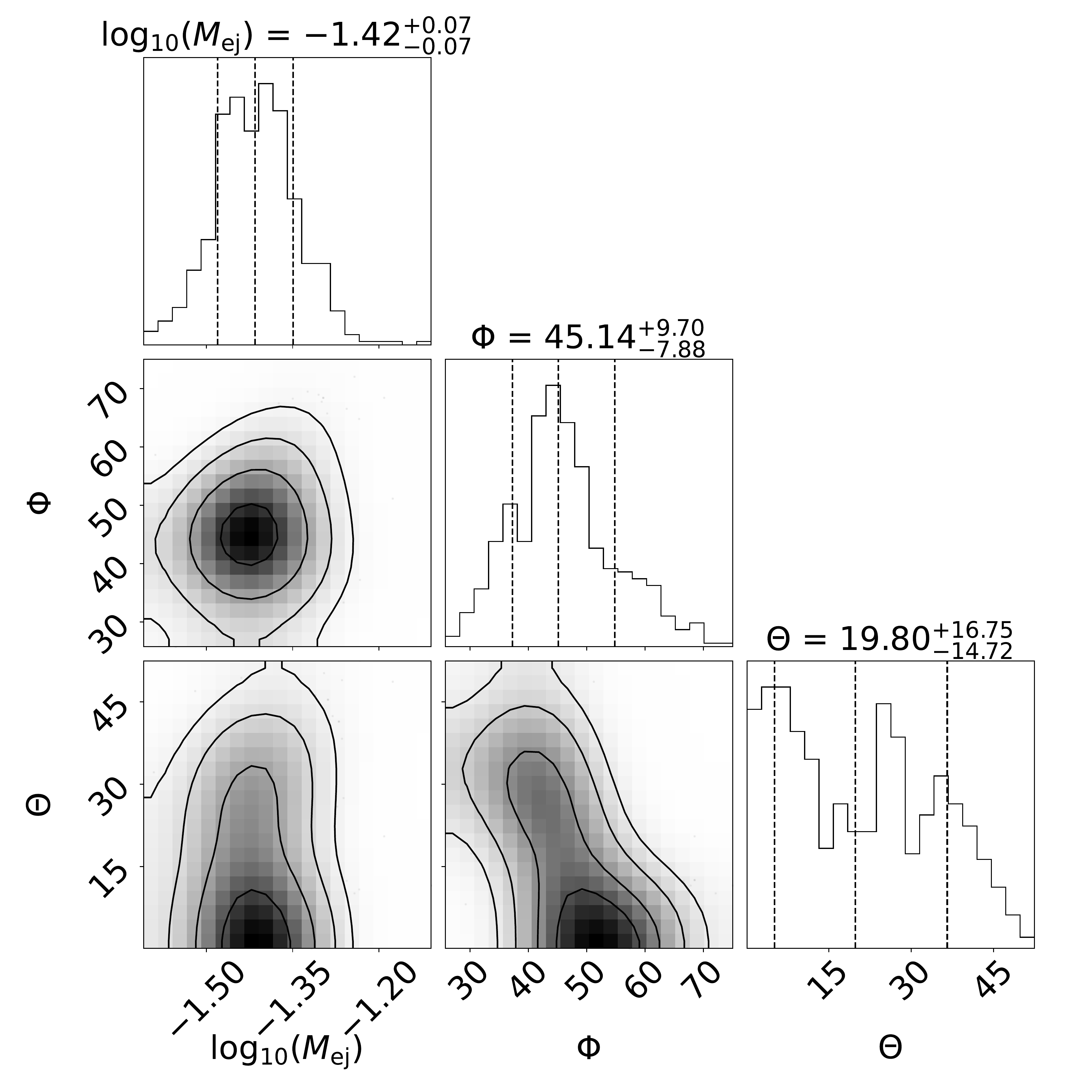}
    \caption{Corner plot showing posteriors of GW170817 for the SSCr model assuming an uncertainty of $\pm 0.5$ mag added in quadrature with the measurement uncertainty for the lightcurve fitting.}
    \label{fig:bullaCorner}
\end{figure}

With the surrogates in hand, we calculate the likelihood as a function of the geometry parameters. In general, we have the ability to allow for a time shift and distance, but we fix these to match the known values for AT2017gfo. Over this $n$-dimensional parameter space (5 dimensional for the SSCr model, 7 for ElDC and ElDCr, and 8 for TorS and TorSr; the number of geometry parameters plus three parameters for inclination, time shift, and distance), we then obtain posteriors for the most likely geometry for each model, and the corresponding lightcurve predicted by the model. In Fig. \ref{fig:GW170817}, we show the maximum likelihood $\chi^2$ fit to the lightcurve data of GW170817, drawn from \cite{CoDi2018b}, using each of the five models, while in Fig. \ref{fig:bullaCorner}, we show an example corner plot for the posteriors on the simulation parameters in the SSCr simulation set.

\begin{figure*}
    \centering
    \includegraphics[width=0.95\linewidth]{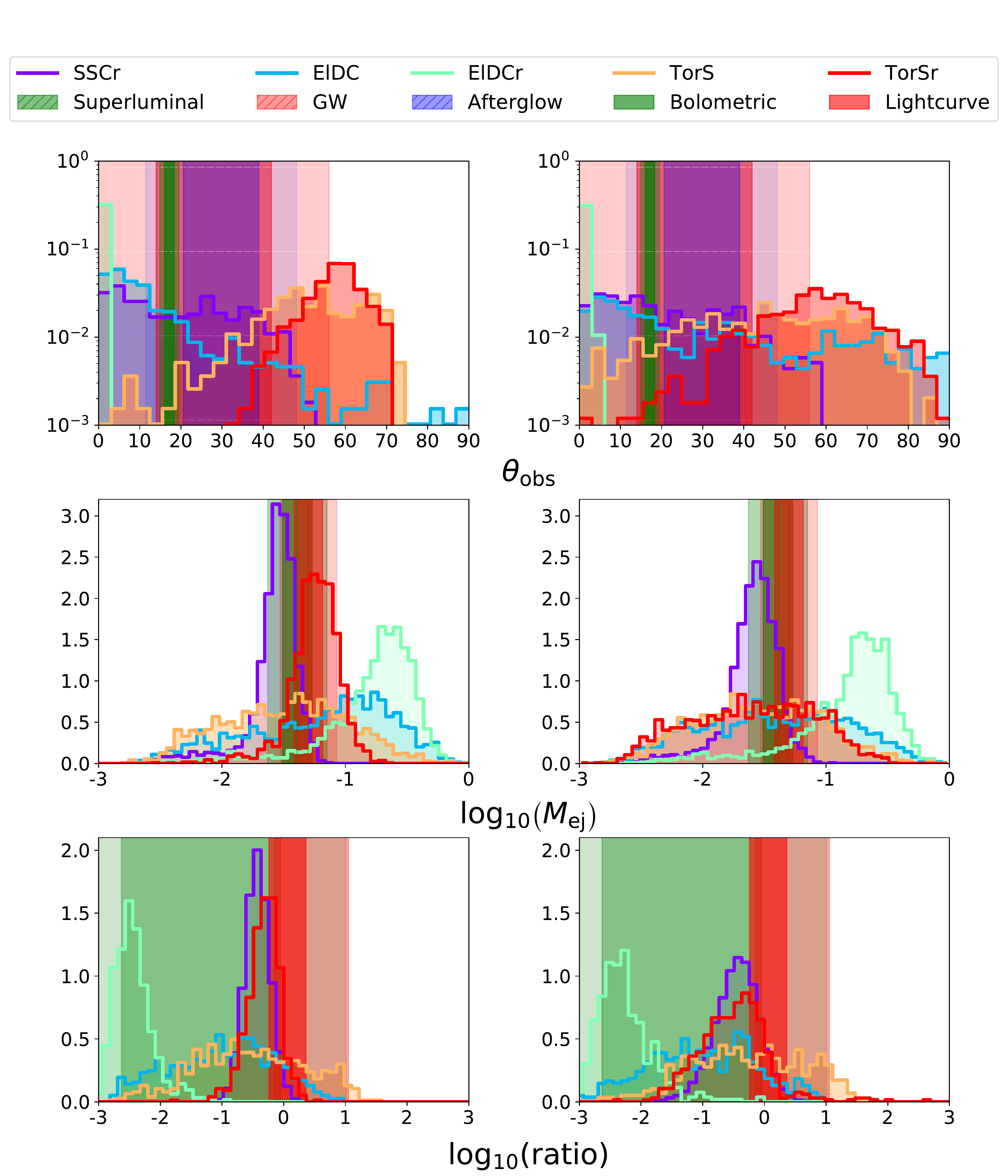}
    \caption{Histograms for the inclination angle (top row), ejecta masses (middle row), and mass ratio (bottom row) constraints based on the five different surrogate models, assuming a $\pm 0.5$ mag (left) and $\pm 1.0$ mag (right) uncertainty for the lightcurve fitting. Dark blue is the SSCr model, light blue is the ElDC model, light green is the ElDCr model, orange is the TorS model, and in red is the TorSr model. In the top row, we also include constraints from gravitational waves  \protect\citep{2017Natur.551...85A}, afterglow modelling  \protect\citep{EeTr2018}, and superluminal motion \protect\cite{HoNa2018}. In the middle and bottom rows, we also include the constraints derived from  \protect\cite{CoDi2018}, which uses our previous method.}
    \label{fig:posteriors}
\end{figure*}

We are interested in comparing the constraints for model parameters represented by the different ejecta geometries. In particular, we compare posteriors for the inclination angle (top row), ejecta mass (middle row) and lanthanide poor to lanthanide rich mass ratio (bottom row) in Fig. \ref{fig:posteriors}.

As a metric for the ``agreement'' or ``overlap'' between the posteriors $f_1, ...,f_n$, we will use the \textit{Matusita} affinity \citep{Matusita1967,Toussaint1974},
\begin{equation}
    A = \int \left[ f_1(x) ... f_n(x)\right]^{1/n} dx.
    \label{eq:agreement}
\end{equation}
In this formalism, if all the posterior density functions are identical, $A=1$, and if any two are absolutely disjoint, $A=0$. The integrand is the geometric mean of the posterior densities, and in general, has the property of being higher if the $n$ posterior densities are closer in value. This functional is close to 1 if the $n$ posteriors are similar in form, and close to 0 if they are different ~\citep[see][for more details]{Matusita1967,Toussaint1974}.

The Matusita affinity is sensitive to the form of each posterior. That is, it not only quantifies consistency in posteriors, but also in form. For example, a constraint of $-2.1 \le \log_{10}(M_{\rm ej}) \le 0.1$ is consistent with a constraint of $-1.3 \le \log_{10}(M_{\rm ej}) \le -1.1$, but there is some notion of an inconsistency in \textit{form}. For our purposes, we want to be able to quantify both these notions of consistency, so this Matusita affinity is appropriate.

We show the posteriors for the inclination angle, total ejecta mass, and ejecta mass ratio in Fig. \ref{fig:posteriors}. On the left hand column of Fig.~\ref{fig:posteriors}, we show the posteriors for the fits when we set the systematic error to $\pm 0.5$ mag, added in quadrature with the measured photometric uncertainty in the samples of AT2017gfo, collected from the data set described in \cite{CoDi2018b}.

Notice that in each of the posteriors, the ElDCr seems to be in disagreement with each of the other posteriors. 
We'll quantify this disagreement by using Eq. \ref{eq:agreement}.
For the inclinations, $A_{\rm inc}=0.01$. However, motivated by the qualitative disagreement from the ElDCr posterior from each of the others, if we replace it with what we used as a prior (simply uniform across $0 \le \theta_{\rm obs} \le 90$), then the affinity increases significantly, to $A_{\rm inc}' = 0.27$ (the prime denotes that we are using the prior instead of the ElDCr posterior). 
Because of the surprisingly narrow posterior for this constraint, we examined the inferred lightcurves for the ElDCr models directly. 
For $\theta_{\rm obs} \notin [0^\circ,3^\circ]$, the lightcurves diverge from AT2017gfo, particularly for late times in the NIR bands. This is consistent with the posterior shown in Fig. \ref{fig:posteriors}.

For the ejecta masses, $A_{\rm mass}=0.41$. Though it is much higher than the inclination affinities, it still suggests systematic differences between the models. Furthermore, the ElDCr model is, once again, the most qualitatively different. When we suppress the ElDCr posterior and replace it with the uniform prior, $A_{\rm mass}' = 0.61$. 

We can also derive constraints for the dynamical to disk wind mass ratio. When we compare the five posteriors, we find the Matusita affinity for the posteriors is $A_{\rm ratio} = 0.10$, and without the ElDCr model, $A_{\rm ratio}' = 0.53$.

This indicates that the ElDCr model yields fundamentally different results from not only the separately simulated ElDC model, but also the SSCr and the TorS and TorSr models. Given the inconstancy with the other models \textit{and} other forms of inclination, ejecta mass and mass ratio measurements \citep{2017Natur.551...85A,EeTr2018,HoNa2018,CoDi2018}, the ElDC model is disfavored as being a geometry that encapsulates AT2017gfo particularly well.

We can also use this method to probe the level of systematic errors within the models. 
To determine the effect of the assumed systematic error on the posteriors, we show posteriors when $\pm 1.0$ mag is added in quadrature with the measurement error on the right hand column of Fig.~\ref{fig:posteriors}.
As expected, the Matusita affinities indicate much more consistent posteriors, with the cost of less constraining posteriors. For the inclinations, $A_{\rm inc}=0.09$, and ignoring the ElDCr model, $A_{\rm inc}'=0.67$. 
This is also the case when considering the ejecta masses, though not quite so dramatically: $A_{\rm mass}=0.57$, and ignoring the ElDCr model, $A_{\rm mass}' = 0.71$. 
For the ejecta mass ratio between the dynamical ejecta and the disk wind ejecta, the affinity again increases significantly when a larger uncertainty is used for the fitting: $A_{\rm ratio} = 0.41$, and without the ElDCr model, $A_{\rm ratio}' = 0.71$. 
Notice that the affinities when we exclude the ElDCr model are generally much closer to the affinities including it.
This indicates that the $\pm 1.0$ mag systematic error budget is generally appropriate, when more specific information about the geometry of the KN is unknown.


\section{Conclusion}
\label{sec:conclusion}

In this article, we have used a combination of kilonova model grids for a variety of ejecta geometries in combination with Gaussian Process Regression to infer source parameters from the AT2017gfo lightcurve. 
Using this method on different geometries and comparing the inferred constraints, we find that there are some discrepancy between the constraints assuming different underlying geometries. 
In particular, we have found that taking into account reprocessing effects gives a much better fit to AT2017gfo; this method has also shown that the resulting inclination constraint from the inclusion of reprocessing can yield constraints inconsistent with those measured by other methods. Indeed, the large disagreement between the ElDCr model's constraint on the inclination angle and the measurements by analyses such as \cite{2017Natur.551...85A,EeTr2018,HoNa2018,Fide2018} implies that this geometry may not be appropriate to describe AT2017gfo. Our method can therefore be used to constrain the form of the geometry of the KNe - when radiative reprocessing is accounted for, the geometry of the KN associated with AT2017gfo could not have been of the form of the ElDC geometry.

We also showed that the Matusita affinity increases dramatically for the inclination angle constraints (and increases a bit more modestly for the ejecta mass and ratio constraints) when an uncertainty of $\pm 1$ mag is used in quadrature with the measurement uncertainty, as opposed to the nominal $0.5$ mag. For this reason, we recommend using at least an uncertainty of $\pm 1$ mag in lightcurve fitting for inclination angle constraints until systematic uncertainties are better under control.

Our method has also shown how to evaluate the efficacy of certain geometries or constrain the geometry of KN ejecta, i.e., it is not necessarily bad that differently constructed but reasonable geometries seem to disagree with one another. In a future work, using a larger range of possible geometries, one could investigate the possibility of more directly constraining the geometry of KN ejecta using a modification of our technique here. 

Altogether, this analysis shows that using more sophisticated and realistic geometry sets, parameterized according to physical variables, should be encouraged. These could be constructed using long term numerical-relativity simulations directly, for example, as opposed to the more common subjective geometrical parameters. In the short term, however, analyses going forward should use more agnostic geometries, i.e., those one that strike the right balance between approximating a KN geometry and being non-committal to any specific features, such as those achieved in the SSCr geometry.


\section*{Acknowledgements}
The authors would like to thank Daniel Kasen and Ryan Wollaeger for making their models publicly available. 
The authors thank Ryan Wollaeger again for his helpful comments and insights into this work.
JH acknowledges support from the University of Minnesota Research Experience for Undergraduates sponsored by REU grant NSF1757388.
M.~W.~Coughlin acknowledges support from the National Science Foundation with grant number PHY-2010970.
TD acknowledges support by the European Union's Horizon 2020 research and innovation program under grant agreement No 749145, BNSmergers. 
NC acknowledges support from the National Science Foundation with grant number PHY-1806990. SA is supported by the CNES Postdoctoral Fellowship at Laboratoire Astroparticle et Cosmologie.

\bibliographystyle{mnras}
\bibliography{references}


\end{document}